\documentclass[%
 reprint,
 amsmath,amssymb,
 aps,
 prapplied,
]{revtex4-2}

\usepackage{graphicx}
\usepackage{dcolumn}
\usepackage{bm}
\usepackage{url}
\usepackage{xcolor}
\usepackage{lineno}

\begin{document}


\title{Nonequilibrium pulse dynamics and metastable latching in nonlinear kinetic inductance detectors}

\author{M. Rouble}
\affiliation{t0.technology, Montreal, QC, H2X 0A4, Canada}
\email{maclean.rouble@t0.technology}

\author{C. Albert}
\affiliation{California Institute of Technology, 1200 California Boulevard, Pasadena, California 91125, USA}

\author{P. Day}
\affiliation{Jet Propulsion Laboratory, California Institute of Technology, 4800 Oak Grove Drive,
Pasadena, California 91109, USA}

\author{H. G. Leduc}
\affiliation{Jet Propulsion Laboratory, California Institute of Technology, 4800 Oak Grove Drive,
Pasadena, California 91109, USA}

\author{M. Dobbs}
\affiliation{Department of Physics and Trottier Space Institute, McGill University, Montreal, QC, H3A 2T8, Canada}

\author{J. Montgomery}
\affiliation{t0.technology, Montreal, QC, H2X 0A4, Canada}


\date{\today}

\begin{abstract}

Microwave kinetic inductance detectors are typically operated at high readout power to raise the detector signal above system noise. At sufficiently large readout power, the current-dependent kinetic inductance couples the detector response to its readout bias. 
Using a nonlinear resonator framework and time-domain circuit calculations, we show that the amplitude, shape, and relaxation time of the driven detector's response depend on both the absorbed energy and on the readout bias.
Strongly driven bias points produce amplified, extended, and non-exponential pulse responses.
Qualitative agreement between calculated and measured pulse responses indicates that these effects are dominated by the driven nonlinear resonator dynamics rather than by altered quasiparticle dynamics.
Beyond resonance bifurcation, sufficiently large pulse events drive the resonator between stable branches, resulting in a metastable latched state which persists after the quasiparticle transient has decayed. 
The pulse energy required for branch switching is set by the readout bias, suggesting a mode of triggered detection with an in-situ tunable threshold. 
Although nonlinear operation requires calibration of the bias- and energy-dependent response, the enhanced pulse amplitude and duration, together with tunable latching and the ability to select these parameters via the readout operating state, are likely to be of interest for single-photon and rare-event experiments, especially those limited by amplifier or system noise.

\end{abstract}

\maketitle

\section{Introduction}

Superconducting resonator detectors such as microwave kinetic inductance detectors (MKIDs) convert absorbed energy into a change in the conductivity of a superconducting film, which presents as a change in the device's resonant frequency and dissipation.\cite{day2003} 
This versatile operating principle is applicable to both continuous power integrating measurements as well as the capture of discrete energy deposition events, enabling MKIDs to support applications ranging from line intensity mapping to transient detection. In a transient detection context, MKIDs can both count individual events and resolve their deposited energies, making them useful for applications including single-photon counting and spectroscopy,\cite{day2024_25um_singlephoton,mazin2019_review_MKIDs_2020s,deRooij2026_thesis}, particle physics and radiation tagging,\cite{cruciani2022_bullkid,mariani2026_muon_veto} or rare-event searches.\cite{cardani2021_calder,temples2024_KIPM_NEXUS}
In this mode, the measured signal is a trajectory in the plane of complex transmission which can be mapped back through the dynamics of quasiparticle production and recombination to infer the energy of the absorbed event. Pulse analysis generally relies on a calibrated response template, with the detected pulse amplitude used as an estimator of the absorbed energy and the rise and decay shapes interpreted in terms of quasiparticle lifetime, phonon dynamics, and resonator bandwidth. 

This template-based picture is well-matched to detector operation in which the response of the detector is locally linear, as the approach implicitly assumes that pulses of different absorbed energy follow approximately the same trajectory through the detector response space, differing primarily by an overall scale factor.
In practice, however, these detectors are often driven with large microwave power to increase measurement sensitivity over system noise sources and to suppress two-level system noise contributions. At sufficiently high readout power, the kinetic inductance becomes appreciably current-dependent. This Duffing-oscillator behaviour is often noted as distortion in the resonator's response to swept-frequency measurements, eventually resulting in resonance bifurcation at sufficiently high readout powers. The onset of bifurcation is often treated as a practical upper limit on the power that can be applied, but previous work \cite{Swenson2013}\cite{rouble2026} has shown that operation in the strongly driven regime can be advantageous. Most of this work, however, has focused on the steady-state frequency-domain response, or small perturbations around a driven operating point.

Feedback between the readout current and the driven resonant frequency enhances responsivity while extending the resonator's relaxation time. These effects have been systematically studied within a two-dimensional parameter space which couples the detector properties to the readout operating state.\cite{rouble2026} Here we apply this framework to the effects of the nonlinear operating state on the detector's response to absorbed energy events of finite scale. Rather than an infinitesimal perturbation about the operating point, a finite pulse drives the system through a larger region of its nonlinear state space. The resulting measured response need not remain a scaled copy of a fixed template: the pulse height, shape, and rise and decay times depend on both the absorbed energy and on the nonlinear operating state.

In some parts of the nonlinear readout operating space, the same resonator can be operated as a threshold detector. Beyond resonance bifurcation, two stable states are accessible, and a sufficiently large injection of quasiparticles can push the system from one to the other. In this regime, the resulting detector response is a binary yes/no response rather than a scaled pulse decay. As we will show, the energy required to trigger this transition is determined by the readout operating state, rather than a fixed property of the detector alone. This suggests a mode of dynamically tunable threshold detection which may be of use in triggered or vetoed acquisition, photon-counting, and other event-based applications. The following sections examine the nonlinear pulse response both below and beyond this latching threshold.

\subsection{Nonlinear operating state}

In a separate work, we define the parameter space for nonlinear operation in terms of a readout generator tone with fixed angular frequency $\omega_g$ which delivers adjustable power $P_g$ to the network.\cite{rouble2026} In this section, we first review this framework and then apply it to pulse dynamics.

The generator's position is defined with respect to the undriven (zero-current) resonant frequency $\omega_0$ by:

\begin{equation}
    x_0 = \dfrac{\omega_g-\omega_0}{\omega_0} \quad{.}
\end{equation}

For non-zero $P_g$, the readout current flowing through the resonance shifts the resonant frequency through the current-dependent kinetic inductance, producing the driven resonant frequency:

\begin{equation}
    \omega_r = \omega_0 + \delta \omega_r
\end{equation}

\noindent such that the normalized detuning of the generator from the driven, shifted resonance is:

\begin{equation}
    x = \dfrac{\omega_g-\omega_r}{\omega_r} \quad{.}
\end{equation}

The resonant frequency shift follows the form of a Kerr nonlinearity: $\delta \omega_r = K n_{ph}$; each photon added to the resonance shifts its resonant frequency by one Kerr coefficient, $K = \frac{-\hbar \omega_0^2}{E_*}$, with $E_*$ setting the scale of the nonlinearity. For a fixed generator frequency, the steady state number of photons in the resonance is related to the applied generator power by: \cite{Anferov2020}

\begin{equation}\label{eq:steady_state_nph_Pg}
    n_{ph} \hbar \omega_g \dfrac{2 Q_c}{\omega_0}  \left[ (\omega_g - \omega_0 - Kn_{ph})^2 + \left(\dfrac{\omega_0}{2 Q_r}\right)^2  \right] = P_g \quad{.}
\end{equation}

Thus $x_0$ and $P_g$ define the experimentally controlled bias point, with $x_0$ defining the range of operating states that are accessible and $P_g$ used to move the system to a chosen point.

For sufficiently large $x_0$, the resonator response bifurcates, with two stable values of $n_{ph}$ (and therefore two values of the nonlinear detuning $x$) corresponding to a single $(x_0, P_g)$ coordinate. The onset of this behaviour is marked by a cusp where

\begin{equation}
    \frac{dP_g}{dn_{ph}} \rightarrow 0 \quad \mathrm{and} \quad    \frac{d^2 P_g}{d n_{ph}^2} \rightarrow 0 \quad{.}
\end{equation}

As the system approaches this point, the relaxation time approaches infinity.
The cusp occurs where:

\begin{equation}\label{eq:foffset_crit}
    x_{0,c} = \dfrac{\sqrt{3}}{2 Q_r} \dfrac{K}{|K|}
\end{equation}

\noindent and

\begin{equation}
    P_{g,c} = \dfrac{2 \hbar \omega_g}{3\sqrt{3} |K| } \dfrac{Q_c \omega_0^2}{Q_r^3} \quad{.}
\end{equation}

For $x_0$ beyond $x_{0,c}$, there is a range of $P_g$ where each value corresponds to two stable states, with the one accessed depending on the hysteresis of the system (i.e. the direction in which the generator power was swept). 
Values of $x_0$ which are within $x_{0,c}$ produce a power-sweep response which is always single-valued and non-hysteretic. However, even in the single valued regime, local feedback between the generator tone and the resonant frequency can still strongly modify the system's responsivity and relaxation time.

At a given nonlinear bias point, the detuning of the driven resonant frequency from the generator is

\begin{equation}
    x = x_0 + \Tilde{E}(x, P_g, Q_i^{-1})
\end{equation}

\noindent where $\Tilde{E} \equiv \frac{E}{E_*}$ is the normalized energy stored in the resonator. Absorbed pair-breaking energy changes both the resonant frequency and the internal quality factor of the resonance, $Q_i$, resulting in the total change to the driven resonant frequency

\begin{equation}\label{eq:nonlinear_optical_responsivity}
    \delta x = \dfrac{\delta x_0 + \frac{\partial \Tilde{E}}{\partial Q_i^{-1}} \delta Q_i^{-1}}{1 - \frac{\partial \Tilde{E}}{\partial x}}
\end{equation}

\noindent assuming the generator power and frequency are fixed. The derivatives are: \cite{Swenson2013}

\begin{equation}
    \dfrac{\partial \Tilde{E}}{\partial x} = \left[ \dfrac{1}{1+x} - \dfrac{8Q_r^2x}{1 + 4Q_r^2 x^2}  \right] \Tilde{E}
\end{equation}

\noindent and 

\begin{equation}\label{eq:E}
    E = \dfrac{2 Q_r^2}{Q_c} \dfrac{1}{1 + 4 Q_r^2 x^2}\dfrac{P_g}{\omega_r} \quad{.}
\end{equation}

The denominator of Eq. \ref{eq:nonlinear_optical_responsivity} describes the strength of the feedback between the readout current and the resonant frequency: when it is greater than 1, the feedback suppresses the response in $x$ to an optical fluctuation (negative feedback); when it is less than one, the response is enhanced (positive feedback). Where it approaches 0 (which occurs only beyond the critical point), the system is unstable and runaway positive feedback drives it to a stable state on the alternate branch. This is seen as bifurcation in swept-frequency measurements at sufficient readout power.

When the system is perturbed about its operating point, the resulting perturbation has a relaxation time $\tau = \frac{-1}{Re(\lambda_{\pm})}$, where 

\begin{equation}\label{eq:linearized_eigenvalues}
    \lambda_{\pm} = -\dfrac{\omega_0}{2 Q_r} \pm \sqrt{(Kn_{ph})^2 - (\omega_0 - \omega_g + 2Kn_{ph})^2}
\end{equation}

\noindent are the eigenvalues of the linearized cavity system.\cite{rouble2026,Eichler2014,YurkeBuks2006,Buks2006}  
At low power, $\tau_{r} = \tau_{r,0} = \frac{2Q_r}{\omega_0}$, the ordinary ring time of the resonator. Together with $\tau_{qp}$, the quasiparticle lifetime, this relaxation time sets the roll-off of the fluctuation spectrum. Near the critical point, $\lambda_+ \rightarrow 0$, and the associated resonator relaxation time becomes sufficiently long to dominate the measured response.

\subsection{Finite transient response}
The preceding discussion describes the steady state of the driven resonator system and the local response to infinitesimal fluctuations about it. This local treatment identifies the properties of the driven system which are dependent on the operating point: the responsivity (via the readout current - resonant frequency feedback factor) and the relaxation time.
An absorbed pulse of energy perturbs the system beyond the local limit, changing the quasiparticle density by a finite amount and moving the system through a trajectory of driven resonator states. Along the excursion, the local feedback factor and relaxation time may vary. The observed response is determined by both the initial operating point from which the trajectory will depart, and by the amplitude of the input pulse.

\begin{figure}[htbp]
    \centering
    \includegraphics[width=\linewidth]{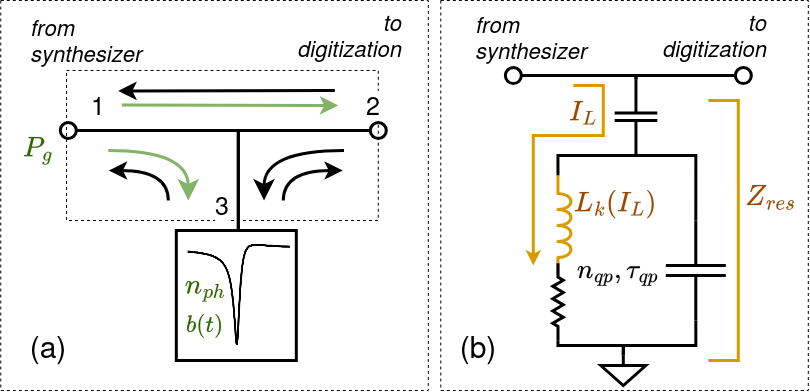}
    \caption{Equivalent descriptions of a driven resonator, with components affected by the readout current highlighted with colours.
    (a) A side-coupled resonator modeled as a one-port Kerr cavity embedded in a symmetric three-port network. A generator tone of power $P_g$ enters the network from the synthesizer, producing an intracavity field $b(t)$ with average photon occupancy $n_{ph}$; the transmitted field is measured at the digitizer. This representation is used to describe the feedback parameter space and infinitesimal perturbations from the steady-state nonlinear detuning. 
    (b) Lumped-element circuit representation of the same resonator. Absorbed optical load changes the quasiparticle density $n_{qp}$, modifying the resonator impedance and undriven resonant frequency, while the readout current $I_L$ modifies the kinetic inductance as $L_k(I_L)$. This representation is used to calculate the time evolution of the resonator system to quasiparticle pulses of finite scale. 
    }
    \label{fig:network_vs_circuit_model}
\end{figure}

To examine the response of the system to an absorbed energy pulse, we combine the local steady state analysis with a lumped-element circuit model of the same resonator, shown schematically in Fig. \ref{fig:network_vs_circuit_model}. The side-coupled Kerr cavity (Fig. \ref{fig:network_vs_circuit_model}a) provides a natural basis for defining the nonlinear operating state, with experimentally controlled quantities $x_0$ and $P_g$ determining the steady-state photon occupation of the cavity, $n_{ph}$, and the nonlinear detuning $x$.
This compact formulation makes transparent the nonlinear operating space and the small-signal dynamics about each operating point, including the critical point, the responsivity and feedback factor, and the relaxation time.

To follow the trajectory of a finite pulse in time, 
it is more convenient to describe the system using a lumped-element circuit model such as in Fig. \ref{fig:network_vs_circuit_model}b.\cite{rouble2025} The resonator is treated a capacitively-coupled tank circuit, with the undriven inductor's impedance (including a reactive and a dissipative component) calculated from the Mattis-Bardeen complex conductivity as a function of quasiparticle number density and lifetime, $n_{qp}$ and $\tau_{qp}$, respectively. The readout current entering the inductor $I_L$ alters this base kinetic inductance as $L_k(I_L)$, providing the nonlinear frequency shift. Changes to the resonator quasiparticle density alter the impedance at the readout frequency ($Z_{res}$) through its zero-current kinetic inductance, $L_k(0)$, and dissipation. This modifies the readout current through the inductor, further altering the total driven kinetic inductance and closing the feedback loop.

The model retains changes to both the reactive and dissipative components of the resonator impedance due to changes in the quasiparticle population. However, we neglect any current-dependent dissipative effects, and assume no influence of the readout current on the quasiparticle population itself.\cite{rouble2026} This approximation, motivated by simplicity in computing the nonlinear effects, is suitable for the resonator measured in this work, for which the reactive component of the readout response is strongly dominant. Sources in the literature report resonators with readout current responses which are primarily reactive, as well as those with significant dissipative components.\cite{Swenson2013,Anferov2020, goldie2013,thomas2020_nonlinear_effects} A more complete analysis should thus include both dissipative and reactive components, as needed.

The choice between the circuit model and Kerr cavity/Lorentzian frameworks is primarily computational: the steady-state response and feedback parameter space are most transparently exposed using the latter, while the time-evolution of the current-dependent inductance is easily computed with the circuit model by injecting a time-varying quasiparticle pulse and tracking the various circuit impedances. The local responsivity and time constant are properties of the nonlinear operating point and do not depend on whether the state is obtained from the circuit model or from the cavity/Lorentzian formulation. 
Combining the two methods allows us to understand the initial point from which a given pulse trajectory will depart, which will determine much of the system's response, as well as the evolution of the trajectory that the driven system then takes.

\section{Nonequilibrium dynamics}\label{sec:nonequilibrium_dynamics}

\subsection{Pulse response}\label{sec:pulse_response}

\begin{figure}[htbp]
    \centering
    \includegraphics[width=\linewidth]{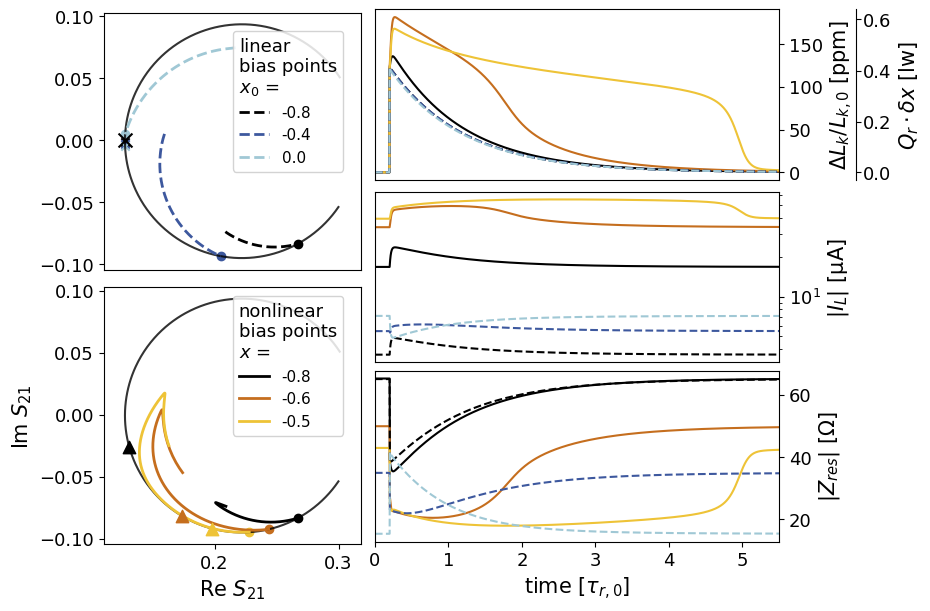}
    \caption{Calculated effect of bias point on pulse response. Identical quasiparticle pulses are applied to the same resonator prepared at different bias points, both linear (blue dashed lines) and nonlinear (orange solid lines). In the linear case, the generator is located at three offset frequencies (where $x_0 \simeq x$ since the nonlinear resonance shift is negligible) from the relaxed resonant frequency, at the same power. An $\times$ marks the location of the undriven resonant frequency. In the nonlinear case, the generator frequency is at fixed $x_0 \simeq x_{0,c}$, with $P_g$ increased to vary the nonlinear detuning, $x$. The location of the driven resonant frequency is marked with a triangle for each detuning on the lower left panel, in the coordinates of the relaxed resonance. Because the resonance is biased in the positive feedback regime, the observed pulse size and relaxation time increase as the resonance is driven closer to the generator, and the response shape increasingly deviates from an exponential. In the linear case, the resonance is unaffected by the generator, so all three bias points produce the same resonant frequency response, as seen in the top right panel. }
    \label{fig:calculated_pulses_various_detuning}
\end{figure}

\begin{figure}[htbp]
    \centering
    \includegraphics[width=\linewidth]{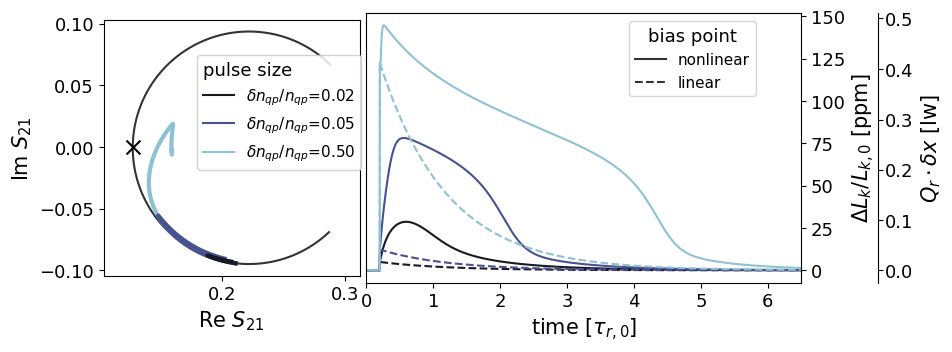}
    \caption{Effect of absorbed pulse energy on pulse response. The resonator is biased to a point just within the critical point, while the size of the injected quasiparticle pulse is varied. The response of the same resonator to the same injected pulses under a linear bias point is shown in dashed lines. In the linear case, the amplitude of the response is proportional to the size of the input pulse and the decay shape follows the same exponential shape as the recombination of the quasiparticles. 
    In the nonlinear case, the response increases with input pulse size but the scales are not linearly related, and the shape of the response is distorted. For small pulses the rise time is notably extended, while for larger pulses the decay is so extended that it appears to approach a shelf before rapidly decaying back to the initial state. 
    The exact shape of the pulse response is determined by the entirety of the path the system traverses through the nonlinear state space, which depends on both the initial operating point and on the energy of the input pulse.
    }
    \label{fig:calculated_pulses_various_sizes}
\end{figure}

We now use the circuit representation shown in Fig. \ref{fig:network_vs_circuit_model} to calculate the detector's response to absorbed energy events of finite scale (Figs. \ref{fig:calculated_pulses_various_detuning} and \ref{fig:calculated_pulses_various_sizes}).
Here, the absorbed energy enters as a time-dependent excess in the quasiparticle density; for simplicity, we consider an instantaneous rise and exponential decay following $n_{qp}^{ pulse}(t) \sim n_{qp} e^{-t/\tau_{qp}}$, with $n_{qp}$ and $\tau_{qp}$ the steady state quasiparticle number density and lifetime respectively. The resulting change in impedance at the generator frequency modifies the readout current flowing through the inductor, which in turn changes the current-dependent kinetic inductance.

When operated in the low-power regime, the system's excursion from the initial steady state is dictated by the production and then recombination of the quasiparticles. Larger absorbed energies produce more quasiparticles, and correspondingly a larger observed response. 
In the nonlinear regime, the response to the pulse is the result of the system's trajectory through the coupled quasiparticle-readout current state space. This results in markedly different responses to pulses of different energies, determined by the nonlinear operating point.

Fig. \ref{fig:calculated_pulses_various_detuning} isolates the impact of operating point on the pulse response by holding the amplitude of the injected quasiparticle pulse constant while varying the readout bias conditions. In the linear low-power case (dashed traces), the bias is varied by placing the generator at different offset frequencies, $x_0$, from the resonance. Since there is effectively zero nonlinear resonant frequency shift at this generator power, $x_0 \simeq x$. At each bias point, the response follows the quasiparticle transient closely, since the current-dependent contribution to the kinetic inductance is negligible. 

In the nonlinear case (solid traces), the generator is fixed near the critical offset, $x_{0,c}$, and $P_g$ is increased to vary $x$. At less negative $x$ (the driven resonator is less detuned from the generator), the larger initial current through the inductor causes the same quasiparticle pulse to produce a larger change in the total kinetic inductance. The nonlinear inductance therefore increasingly dominates the resonator response.

The readout current is modulated by the total impedance at the generator frequency. This includes contributions from both the underlying zero-current impedance and the nonlinear contribution from the current itself. As seen in the lower right panel of Fig. \ref{fig:calculated_pulses_various_detuning}, at highly nonlinear biases, the impedance reaches a minimum (and the current a maximum) at a significant delay after the peak of the quasiparticle transient itself. This causes the extended decay seen in the resonant frequency shift, as a significant fraction of the total kinetic inductance is due to the current, extending the response beyond the timescale of the quasiparticle recombination.

Fig. \ref{fig:calculated_pulses_various_sizes} shows the complementary case, where the operating point fixed while the amplitude of the injected quasiparticle pulse is varied. 
The operating point is highly nonlinear but below the critical point in both $x_0$ and $P_g$. The response is large and distorted, with an extended rise and fall time. The shape of the response varies significantly as a function of the input pulse amplitude because the system traverses different parts of the nonlinear state space. As it does so, both the time constant and the responsivity of the driven system vary widely, shaping the response. The response of the same resonator to the same input pulses, but under a linear bias condition, is shown as dashed lines. These scale with the input pulse amplitude and retain their exponential decay shape.

\begin{figure}[htbp]
    \centering
    \includegraphics[width=\linewidth]{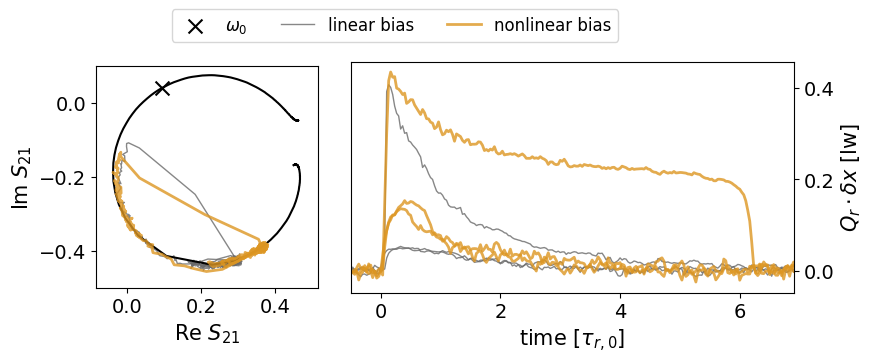}
    \caption{Response of a resonator to absorbed photons, measured in the lab, at a linear (thin grey lines) and nonlinear (thick orange lines) operating point. The data are shown on the IQ plane (left, overlaid on a low-power frequency sweep of the same resonator) with the original timestreams shown at right. Although the readout current is non-zero in both cases, the linear bias point is at a substantially less negative $x_0$ and smaller $P_g$ than the nonlinear bias point. At the linear bias point, the resonator response follows an exponential decay, with a peak amplitude that scales with the energy of the absorbed pulse because the response is dominated by the quasiparticle transient. At the nonlinear bias point, the response is dominated by the current-dependent inductance, and thus is distorted and extended in agreement with the calculations in Figs. \ref{fig:calculated_pulses_various_detuning} and \ref{fig:calculated_pulses_various_sizes}. Under both the linear and nonlinear bias conditions, larger pulses (such as these likely corresponding to cosmic ray strikes) produce larger responses, but at the nonlinear bias point, larger pulses also produce a response which is greatly elongated, lasting many times the linear decay time, $\tau_{r,0}$, and producing a distinctive shelf-like shape before decaying back to the starting point. This distortion is not seen in the linear bias condition, regardless of the size of the input pulse event.
    }
    \label{fig:measured_pulses_nonlinear_and_linear}
\end{figure}

Our calculations in Figs. \ref{fig:calculated_pulses_various_detuning} and \ref{fig:calculated_pulses_various_sizes} provide a useful interpretation of the measured pulse responses shown in Fig. \ref{fig:measured_pulses_nonlinear_and_linear}. 
The measured pulse response was obtained by illuminating a far-IR sensitive MKID with a bandpass-filtered IR source. The MKID is made from a niobium interdigitated capacitator and a low-volume aluminum inductor that forms a resonant structure tuned for direct absorption at 25~$\mu$m. This device is thoroughly characterized in \cite{day2024_25um_singlephoton}. The source is a broadband black-body emitter made from black Acktar Metal Velvet foil epoxied to a copper plate. Several stages of filtering along the optical path reduce both the total power and the bandwidth incident on the detector. The filter stack is centered at 25~$\mu$m and has sub-percent transmission outside of the 21-29~$\mu$m. A diagram of the optical setup and the shape of the filter spectral response may be found in \cite{day2024_25um_singlephoton}. 
This results in the absorption of discrete single-photon events, spanning a narrow (but not monochromatic) range of energies.

The resonator is biased at two points: one at high power and $x_0$ near but within the critical value (orange traces in Fig. \ref{fig:measured_pulses_nonlinear_and_linear}; hereafter referred to as the nonlinear bias point), and one further from the critical point in both $P_g$ and $x_0$ (thin grey traces in Fig. \ref{fig:measured_pulses_nonlinear_and_linear}; the linear bias point).
At the linear bias point, pulses of different energies retain an exponential shape and scale with the amplitude of the absorbed pulse. This is consistent with the response being dominated by the quasiparticle transient. At the nonlinear bias point, the response is strongly energy-dependent but is no longer a simply scaled exponential decay. Small events show a rounded peak and smooth, slightly extended decay, while larger events produce an extended shelf before relaxing back to the initial state. The extended decay time may be many times the linear decay time, $\tau_{r,0}$ (here extracted by fitting an exponential to the measured pulses at the linear bias point). This qualitative behaviour matches that predicted in Figs. \ref{fig:calculated_pulses_various_detuning} and \ref{fig:calculated_pulses_various_sizes}.

We emphasize the distinction between the quasiparticle decay and the decay of the measured resonator response. In the calculations in Figs. \ref{fig:calculated_pulses_various_detuning} and \ref{fig:calculated_pulses_various_sizes}, the input quasiparticle pulse shape is fixed by construction. \textbf{The observed extended response is therefore due to the nonlinear inductance contributed by the readout current, without requiring any extension of the quasiparticle lifetime.} The qualitative agreement between the measured and calculated responses suggests that a decay time extracted from such a pulse would represent the decay of the driven detector system, rather than the quasiparticle lifetime alone.

For pulse-detection applications, this behaviour is likely to be both useful and a complicating factor. 
The choice of nonlinear operating point can substantially increase the amplitude and duration of the response to a given event, improving sensitivity over amplifier and other additive system noise. However, the response is no longer described by a fixed template scaled by the absorbed energy. A calibration of nonlinear pulse response therefore must account for both the readout bias condition and the pulse size.

\subsection{Latching}\label{sec:latching}

The extended shelf-like pulse responses discussed in Sec. \ref{sec:pulse_response}, while long compared with the undriven resonator relaxation time, are return trajectories: the resonator eventually relaxes back to its initial driven state. In contrast, under certain nonlinear operating conditions, the detector response may not return to its initial state at all. In this case, the detector may be considered to have `latched.' This is distinct from simply a very slow decay, and occurs when a quasiparticle pulse drives the system onto a different stable branch of the nonlinear state space. 
Once in the driven metastable state, it will remain there even after the quasiparticle transient itself has decayed completely. In this case, the quasiparticle pulse acts as a trigger, and the detector's response becomes a yes/no output.

Branch switching is possible only when the nonlinear operating point admits more than one stable driven state. This condition can be met for generator frequencies at and beyond the critical offset $x_{0,c}$, for generator powers for which the derivative of Eq. \ref{eq:steady_state_nph_Pg} at that frequency (here expressed in terms of the nonlinear detuning) $\left.\frac{dP_g}{d x} \right|_{x_0} \rightarrow 0 $. When this condition is met, there is a region in detuning where the feedback factor $1 - \frac{\partial \Tilde{E}}{\partial x}$ (Eq. \ref{eq:nonlinear_optical_responsivity}) goes to zero. This corresponds to a region of runaway positive feedback. If the pulse deposits enough energy in the resonator to push it into the runaway feedback regime, it will cross the unstable region to the other stable branch, where it will remain.

\begin{figure}[htbp]
    \centering
    \includegraphics[width=\linewidth]{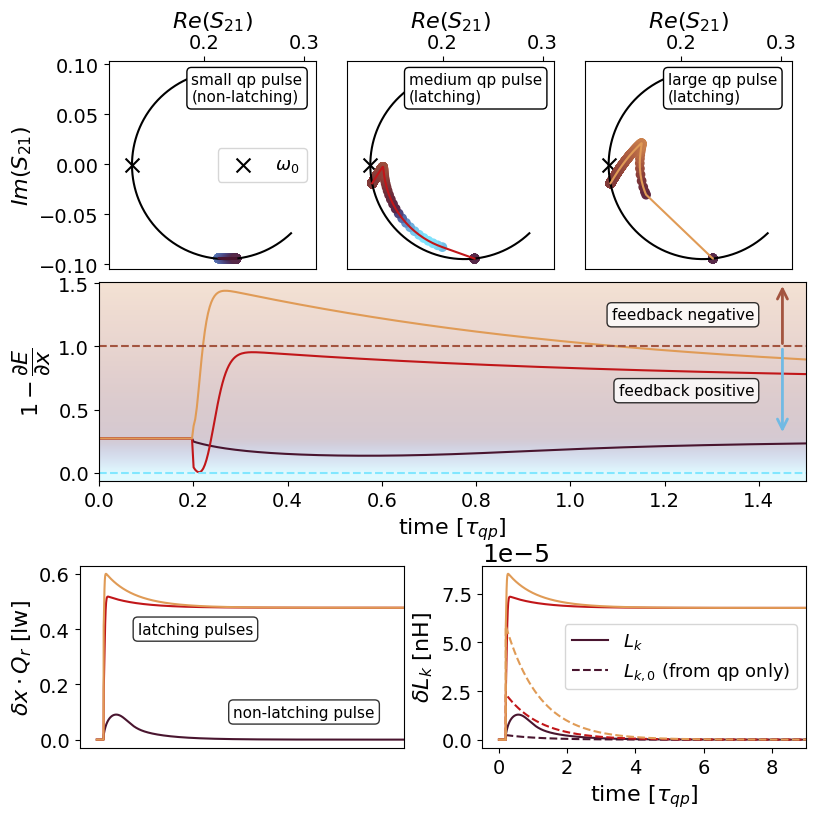}
    \caption{
    Calculated response of a resonator biased beyond the critical point to injected quasiparticle pulses of different sizes. 
    The smallest input pulse perturbs the resonator but does not drive it into the runaway feedback region, so it relaxes to its initial state. Larger pulses enter or cross the runaway feedback region (highlighted in bright blue), switching the system to the coordinate on the alternate stable branch corresponding to the same operating point $(x_0, P_g)$. 
    The upper panels show the trajectories of the resonator's response in the complex transmission plane. The middle panel shows the evolution of the feedback factor along each trajectory (also encoded on the complex plane by point colour). The lower panels compare the quasiparticle-induced perturbation with the resulting nonlinear resonator response in units of resonant frequency and kinetic inductance change. Although the amplitudes of the triggering pulses differ, the latched state is dependent only on the initial operating point, and thus is the same between the two latching pulse trajectories.
    }
    \label{fig:calculated_latching_pulses}
\end{figure}

This is illustrated in Fig. \ref{fig:calculated_latching_pulses}. With the resonator biased at $x_0$ beyond the critical offset, pulses of different energies are injected and the response calculated. The top row of panels shows the pulse trajectories on the IQ plane, superimposed on a low-power frequency sweep of the resonator. The value of the feedback factor at each point in the trajectory is indicated by the colours of the points on the IQ plane. These correspond to the middle panel, which shows the pulse trajectories through the feedback factor as a function of time. The lower panels show the change in the nonlinear detuning throughout the response (left) and the change in the kinetic inductance (right), broken down into its zero-current value (contributed by the quasiparticle transient alone) and the total including the contribution from the readout current.

Although at the chosen bias point latching is possible, the smallest pulse (top left panel) has insufficient energy to drive the resonator into the unstable region, and thus it returns to its initial state, although with evident distortion and slowing from the nonlinear operating point.
The two larger input pulses cause latching. The middle-energy pulse is small compared with the total system response, as it provides just enough energy to nudge the resonator into the unstable regime (bright blue) before decaying. The driven resonator system then crosses the unstable region and settles on the other stable branch.
The largest input pulse (likely a cosmic ray strike) deposits sufficient energy that, after the instantaneous rise in $n_{qp}$, the resonator has already crossed the unstable region. The system then follows the quasiparticle transient and decays exponentially to the new stable state.
It is noteworthy that although the energy deposited by these pulses differs by an order of magnitude, the final state of the system is identical. This is because the final state corresponds to the same generator power on the other stable branch, where the resonator impedance is sustained by the nonlinear inductance from the readout current.

\begin{figure}[htbp]
    \centering
    \includegraphics[width=\linewidth]{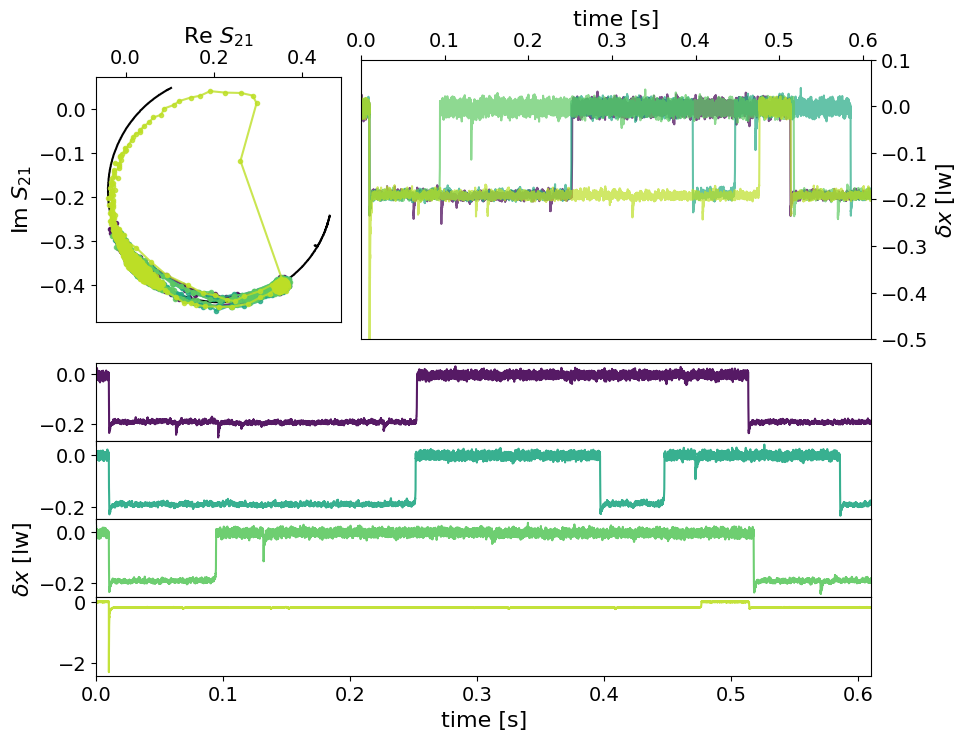}
    \caption{Measured latching response. The same resonator as in Fig. \ref{fig:measured_pulses_nonlinear_and_linear} is illuminated with IR photons while biased slightly beyond the critical point. Upper panels show four measurements of the response under fixed operating conditions, in the plane of complex transmission and as continuous time streams. Sufficiently large absorbed energy events trigger the resonator to switch from its initial driven state to a metastable latched state corresponding to a stable solution on the other branch of the resonator transfer function, where it remains after the quasiparticle transient has decayed. The displacement of the latched state is determined by the generator power at the operating point, and thus is identical between the measurements shown regardless of the triggering pulse amplitude. Return to the initial state occurs due to fluctuations (reductions) in the resonator quasiparticle population or amplitude noise on the generator.  }
    \label{fig:measured_latching_pulses}
\end{figure}

Fig. \ref{fig:measured_latching_pulses} shows observations of latching behaviour measured using the same resonator featured in Fig. \ref{fig:measured_pulses_nonlinear_and_linear}. The measurement setup is the same, but the detector's readout operating point is now slightly beyond the critical point. Under these conditions, sufficiently large pulses trigger the system to latch to the alternate stable branch, while smaller pulses return to the initial state. The figure compares four approximately half-second timestream sample of the resonator at this operating point. During each capture, the resonator latches and unlatches multiple times, remaining in each state for many multiples of the undriven resonator relaxation time. 
The upper panels display the events as measured on the IQ plane (left) and in time (right; with a truncated y-axis to show detail), while the lower panels show the full scale of each measurement individually. While the size of the initial pulse trigger varies between the latching events, the system settles to the identical metastable state once the quasiparticle transient has decayed. This corresponds to the alternate stable operating point for the given generator power and frequency (which is the same between all four measurements).

Unlike in a regular pulse decay where the system smoothly relaxes to its initial state, returning from the latched state requires crossing the separatrix in the opposite direction. In the measurements of Fig. \ref{fig:measured_latching_pulses}, this occurs due to noise-induced fluctuations, rather than an intentional reset. These fluctuations may be random reductions in the resonator quasiparticle population, or amplitude noise on the generator itself.

It is interesting to note also that the system remains sensitive to pulses while in the latched state. As can be seen in the middle panel of Fig. \ref{fig:calculated_latching_pulses}, the latched system is still in the positive feedback portion of the resonance bandwidth, but is at a much lower loopgain (larger value of the feedback factor, $1 - \frac{\partial \Tilde{E}}{\partial x}$) than the initial state. 
The response to pulses when the resonator is in this state is therefore less strongly enhanced relative to that when the system is in the initial unlatched state. Correspondingly, the amplitude of the resonator noise in the latched state is visibly reduced relative to the noise in the unlatched state.

Once the operating generator frequency offset $x_0$ is within the region for which an alternate stable state exists, pulses of sufficient energy will cause the system to latch. The threshold for this energy is, at fixed $x_0$, defined by how far from the runaway feedback region the system is initialized. This is defined by the applied generator power (and consequently the nonlinear detuning from the driven resonant frequency).

Thus, the energy threshold to latch is not a fixed property of the resonator, but is tunable through the generator frequency and power. 
\textbf{Given suitable calibration of a detector's response, this in principle would allow an operator to both select and dynamically alter the energy threshold for event detection, which may be of interest for photon-counting or other rare-event searches.} Such an operating mode trades linear pulse-height information for a large persistent state change, and would require detailed characterization of the energy threshold as a function of bias conditions, as well as the conditions for reset and rates of noise-triggered switching.

\section{Conclusion}

We have examined the response of a nonlinear kinetic inductance detector to absorbed energy events of finite scale. Within a parameter space defined by a readout generator frequency offset from the undriven resonant frequency and an applied readout power, we show that the choice of nonlinear operating state 
affords control over both the amplitude and duration of the detector's response.

Feedback effects between the readout tone and the resonant frequency through the current-dependent kinetic inductance produce an enlarged and non-exponential pulse response, with amplitudes that do not linearly scale with the input pulse energy. For operating points beyond the critical condition, these dynamics produce metastable latching between states on alternate branches of the resonator response curve, with an input energy threshold that is set by the readout state.

These effects may be advantageous in applications limited by amplifier or other additive system noise sources, as the increased amplitude and duration of the response may improve detectability. Beyond bifurcation, the readout operating state additionally provides an in-situ tunable threshold for triggered detection. These advantages come at the cost of reduced simplicity in calibration, as the measured pulse height no longer maps directly to absorbed energy. In the latching regime, this requires characterization of the latching threshold energy, reset conditions, and susceptibility to spurious transitions triggered by noise.

\section{Acknowledgments}

The McGill authors acknowledge funding from the Natural Sciences and Engineering Research Council of Canada and the Canada Research Chairs Program. C. Albert was supported by the National Aeronautics and Space Administration (NASA) Space Technology Mission Directorate (STMD) through the NASA Space Technology Graduate Research Opportunities (NSTGRO) Fellowship under grant number 80NSSC24K1395 (PI: J. Zmuidzinas).

\bibliography{bibliography}

\end{document}